\documentstyle[preprint,aps]{revtex}
\begin{document}
\draft
\preprint{}
\title{
Integrability and ideal conductance at finite temperatures}
\author{H. Castella$^{(1,2)}$, X. Zotos$^{(1)}$}
\address{
 (1) Institut Romand de Recherche Num\'erique en Physique des
Mat\'eriaux (IRRMA), \\
PHB-Ecublens, CH-1015 Lausanne, Switzerland\\
 (2) D\'epartement de Physique de la Mati\`ere Condens\'ee,\\
24,quai E. Ansermet, CH-1211 Gen\`eve, Switzerland}
\author{P. Prelov\v sek}
\address{J. Stefan Institute, University of Ljubljana, \\
61111 Ljubljana, Slovenia}
\date{Received\ \ \ \ \ \ \ \ \ \ \ }
\bigskip\bigskip
\maketitle
\begin{abstract}
We analyse the finite temperature charge stiffness $D(T>0)$, by a
generalization of
Kohn's method, for the problem of a particle interacting with a fermionic
bath in one dimension. We present analytical evidence, using the Bethe
ansatz method,
that $D(T>0)$ is finite in the integrable
case where the mass of the particle equals the mass of the fermions and
numerical evidence that it vanishes in the nonintegrable one of
unequal masses.
We conjecture that a finite $D(T>0)$ is a generic property of
integrable systems.
\end{abstract}



This work relates the finite temperature charge transport
to the response of energy levels to an infinitesimal flux and proposes a
connection to the integrability of quantum systems.

Starting with the formulation by Kohn \cite{Kohn}, the charge stiffness
(or Drude weight)
representing the weight of the delta function contribution $D\delta(\omega)$
part in the dynamical conductivity $\sigma(\omega)$,  has been
investigated as a criterion for a metallic, superconducting or
insulating state \cite{Suth,Scal}.
The original approach \cite{Kohn} relates $D$ to
the response of the ground state energy to a magnetic flux  and thus
requires only the calculation of an equilibrium  property, bypassing
a complete evaluation of the Kubo formula\cite{Kubo}.

A similar concept appears also in the study of the conductance of
disordered metallic systems. Starting with the work of
Thouless\cite{Thoul} a close relation has also being
established between the conductance, the sensitivity on boundary
conditions (being equivalent to the introduction of a flux) and the statistical
properties of {\it single-particle} energy spectra \cite{w} of metallic
systems.

Recently it has also been observed that the level statistics in
{\it many-particle} correlated systems is closely related to the integrability
of the system\cite{Bell,DiSZ}. Still the implications of this fact on transport
quantities in correlated fermion systems at finite temperatures $T>0$,
e.g. a nonzero d.c. resistivity
$\rho(T)>0$ or on the contrary a possible finite charge stiffness $D(T)>0$,
have not been known so far.

In this work we analyse the relation between integrability and transport
in the generic problem of
a single tagged particle moving in a bath of fermions by a
generalization of Kohn's method at finite temperatures.

{\it Kohn's method at} $T > 0$:
We consider a general tight-binding Hamiltonian of the form
\begin{equation}
\hat H=-t\sum_i( e^{i\phi} d^{\dagger}_{i+1} d_{i} + H.c.)+\hat H_{int}=
\hat T + \hat H_{int}, \label{eq1}
\end{equation}
representing a 1D system of length $L$ with periodic boundary
conditions pierced by a flux  $L \phi$, using the Peierls construction.
$\hat T$ is the kinetic and $\hat H_{int}$ the interaction part
of the Hamiltonian.
{}From the Kubo formula\cite{mald,Suth} we can relate the imaginary part
of the dynamical conductivity $\sigma''(\omega)$
to the charge stiffness $D$,
\begin{equation}
D= \frac{1}{2}[\omega \sigma''(\omega)]_{\omega \to 0}
=\frac{1}{L} \{ \frac{1}{2} \langle  -\hat T \rangle -
\sum_{m\neq n} p_n \frac{ \mid \langle n \mid \hat j \mid m \rangle \mid ^2}
{\epsilon_m - \epsilon_n} \}, \label{eq2}
\end{equation}

\noindent
where $\hat j$ is the current operator,
$\langle \hat T\rangle$ the thermal expectation value of the kinetic
energy and $p_n=e^{-\epsilon_n/T}/Z$ the Boltzmann weight for an
eigenstate $\mid n\rangle$ of Hamiltonian (\ref{eq1}) with energy
$\epsilon_n$.

On the other hand\cite{Kohn} we can evaluate, by second order
perturbation theory, for $\phi \to 0$ a shift of the level
$\mid n\rangle$:
\begin{equation}
\epsilon_n(\phi)= \langle n\mid \hat H(\phi=0)\mid n\rangle
-\phi \langle n \mid \hat j \mid n \rangle
-\phi^2 \sum_{m\neq n} \frac{ \mid \langle n \mid \hat j \mid m \rangle
\mid ^2}{\epsilon_m - \epsilon_n} -
\frac{1}{2} \phi^2 \langle n \mid \hat T \mid n \rangle .  \label{eq3}
\end{equation}

Extracting second order terms in $\phi$ (the curvature of levels)
we see that:
\begin{equation}
D =
\frac{1}{L} \sum_n p_n D_n=
\frac{1}{L} \sum_n p_n
\frac{1}{2} \frac {\partial ^2 \epsilon_n (\phi) }{\partial \phi^2} .
\label{eq5}
\end{equation}

In this demonstration we assumed that the levels $\mid n\rangle$
are nondegenerate but  the same procedure can be used for degenerate levels
after applying first order degenerate perturbation theory to lift the level
degeneracy.
This expression for $D(T)$ generalizes Kohn's method to $T>0$ and reduces
its evaluation to the calculation of curvature of levels
under an external flux.
Easier than the complete evaluation of Kubo formula, it can be
performed either numericaly on finite size systems followed by finite
size scaling or in some cases (e.g. integrable systems)
analytically\cite{peter}.
It also provides a basis for analysing the conditions for the occurence
of ideal conductance.

The ground state of the metal should be characterized by $D_0=D(T=0)>0$
\cite{Kohn,Suth,Scal},  where $D_0$  is a measure of
of the charge carriers {\it coherent} motion.
On physical grounds it is commonly assumed \cite{Scal} that
$D(T>0)=0$, for generic
macroscopic interacting systems (involving Umklapp scattering),
except in the  superconducting phase.  This statement, never really
proved to our knowledge for interacting fermions, implies according
to our Eqs.(\ref{eq2},\ref{eq5})
that the positive curvature of the ground state $D_0$ is
cancelled by (mostly) negative curvatures of low lying
states, a situation reminiscent of the mechanism for phase transitions and
the dissapearance of long range order.

{\it Model system:}
The Hamiltonian describing a particle in a bath of spinless fermions is:

\begin{equation}
\hat H=-t_h\sum_i (e^{i\phi} d^{\dagger}_{i+1} d_i +H.c.)
-t\sum_i (c^{\dagger}_{i+1} c_i + H.c. ) +
U\sum_i d^{\dagger}_i d_i c^{\dagger}_i c_i ,\label{eq6}
\end{equation}
where  $c_i ( c^{\dagger}_i )$ are annihilation (creation) operators for
$N$ spinless fermions and $d_i ( d^{\dagger}_i )$ for the (tagged) particle
on an $L$ site chain with periodic boundary conditions. The interaction
comes only through the on-site repulsion $U>0$. The current $\hat j$
in (\ref{eq2}) refers to the particle only (
$\hat j=-it_h\sum_i d^{\dagger}_{i+1} d_i + H.c $)
and $[\hat j,\hat H]\neq 0$. The volume normalization $1/L$ in
(\ref{eq5}) is absent as the current refers only to one particle.

The model (\ref{eq6}) is well adapted to our study:
it is integrable by the Bethe ansatz method in the case of equal masses, i.e.
$t_h = t$ \cite{mcg} (equivalent to the problem of a Hubbard
chain in a nearly polarized state $S^z=S_{max}^z-1$) and
nonintegrable for unequal masses ($t_h\neq t$).
We have also previously studied $D_0$\cite{xz} of the particle
as well as its quasiparticle properties\cite{cz}.

We first present the results for $D(T)$
from exact numerical diagonalization of the Hamiltonian (\ref{eq6})
on finite size systems. We will consider only the $N=L/2$ case, which in this
system should be analogous to all other fillings.
By numerical calculation of the energy spectrum
with and without a small flux $\phi$ (typically $\phi \simeq 10^{-4}$) we
deduce the energy level curvature from:
\begin{equation}
D_n=\frac{1}{2}
\frac{\epsilon_n(\phi)+\epsilon_n(-\phi)-2\epsilon(0)}{\phi^2} .\label{eq7}
\end{equation}

In Fig.1 we present $D(T)$ as a function of $t_h/t$,
for different temperatures $T$, size systems (from 6 to
14 sites) but fixed $U/t=2$.
We observe: {\it a)} a nonmonotonic behavior of $D(T)$ as we sweep
through the integrable $t_h=t$ point for $T>0$,
{\it b)} a very weak dependence of $D(T)$
with system size for the integrable case $t_h=t$
while a rather strong one for the nonintegrable ones.

These observations are the first hints that something particular happens
at the integrable point. To further study this behavior, we present
in Fig.2  $D(T)/D_0$ as a function of $1/L$ for $t_h=t$.
We indeed observe a scaling in $1/L$ which we also know that it is the relevant
one to recover the ground state $D_0$\cite{xz}.
The circle at $1/L=0$ indicates the result of
the Bethe ansatz calculation for $t_h=t$ in very good agreement with
the extapolated numerical results.

The same plot in Fig.2 of $\ln D(T)/D_0$ as a
function of $L^2$ for $t_h=0.5, 1.2, 2.0t$ seems to indicate a
strong $L$ dependence following a law
$D(T)\simeq e^{-\gamma L^2}$ for these nonintegrable cases.
This is consistent with a rapid
decrease of $D(T)$ for systems larger than the mean free path.

Next to obtain an impression of the overall behavior of $D(T)$ as function
of temperature we present in Fig.3 $D(T)/D_0$ for $t_h=t, 0.5t$
and different size systems. For $t_h=0.5t$ $D(T)/D_0$ seems to
scale to zero for any $T>0$ as we found above while for $t_h=t$
there is a smooth decrease with T.
At high temperatures $D(T) \propto 1/T$, vanishing for $T\to \infty$
a general feature of systems on a lattice. Finally the discrepancy from the
free particle $D(T)$ indicates that the temperature
effects cannot be accounted for by a simple scaling of $D_0$.

{\it Bethe ansatz analysis:}
We present here an analytical approach for the calculation of $D(T)$
for the integrable case using the Bethe ansatz method
along the line of reference \cite{xz}. The starting point is
the Lieb-Wu\cite{lw} solution of the Hubbard model adapted for one spin up
fermion (the tagged particle) and $N$ spin down fermions.
The Bethe ansatz
wavefunctions, in the presence of flux $\phi$,
are then characterized by $N+1$ quantum numbers $k_j$ given
by the following equations (we take $t_h=t=1$):
\begin{equation}
Lk_j=2\pi I_j+\theta(\sin k_j-\Lambda),~~~j=1,\ldots,N+1 \label{eq8}
\end{equation}
\begin{equation}
\theta(p)=-2\tan^{-1}(4p/U) \label{eq9}
\end{equation}
\begin{equation}
L\sum_{j=1}^{N+1}k_j=2\pi \sum_{j=1}^{N+1}I_j+2\pi J + L\phi .
\label{eq10}
\end{equation}

Every state is characterized by a set of half-odd integers $\{I_j\}$
and the (half-odd) integer $J$ for (even) odd number of fermions.
However this set of states(regular)
does not constitute a complete set; to obtain a
complete set we must include states
representing bound states of energy of order U. The simplest way to include
these states is by considering an electron-hole transformation
($\tilde c_i=c^{\dagger}_i$, $\tilde d_i=(-1)^i d_i$) and then
solving Eq.(\ref{eq8}) but for a $L-N$ number of fermions
(equal to the number of holes).

The total energy of a regular state is given by :
\begin{equation}
E=\sum_{j=1}^{N+1} \epsilon(k_j)=\sum_j (-2\cos k_j) ,\label{eq11}
\end{equation}
\noindent
and for a bound state:
\begin{equation}
E=U+\sum_{j=1}^{L-N+1} 2\cos k_j .\label{eq12}
\end{equation}

In the following we present explicitely only the analysis for the
contribution of regular states as it is clear how to include the bound
ones by the prescription described above\cite{fut} (although we include
both contributions in the results presented in Fig. 2 and 3).

To order $1/L$ and for $UL\gg 1$ the values of $k_j$ are given by:
\begin{equation}
k_j=k_j^0+\frac{1}{L}\theta(\sin k_j^0-\Lambda),
{}~~~k_j^0=\frac{2\pi I_j}{L} .\label{eq13}
\end{equation}

To order one then the energy and momentum of a state can be written as:
\begin{equation}
E=\sum_j \epsilon(k_j^0)+\frac{2}{L}\sin k_j^0~
\theta(\sin k_j^0-\Lambda) \label{eq14}
\end{equation}
\begin{equation}
\frac{1}{L}\sum_j \theta(\sin k_j^0-\Lambda)=
\frac{2\pi J}{L} +\phi .\label{eq15}
\end{equation}

In this scheme the interaction between the particle and the fermions is
represented
through second term in Eq.(\ref{eq14}), a correlation energy;
the coupling to the flux $\phi$ is through the collective
coordinate $\Lambda$. In the thermodynamic limit, defining a density
$\rho(k)$ $(-\pi \leq k \leq +\pi)$ we obtain:
\begin{equation}
E(\rho(k),\Lambda)=\frac{L}{2\pi}\int dk \rho(k)
( -2\cos k+\frac{2}{L}\sin k~ \theta(\sin k-\Lambda) )\label{eq16}
\end{equation}
\begin{equation}
\frac{1}{2\pi} \int dk \rho(k) \theta(\sin k-\Lambda)=
P+\phi,~~~ P=\frac{2\pi J}{L} .\label{eq17}
\end{equation}

To calculate the thermal average of the curvature of energy
levels we will work in the grand canonical ensemble for the system
plus the one particle. Assuming that the equilibrium distribution of $k's$
is not affected by the presence of the one particle we use the Fermi-Dirac
distribution for free fermions $f(k)=1/(1+\exp( (\mu-\epsilon(k))/T)$
($\mu$ is the chemical potential). We then assume that
the distribution of $\Lambda's$ is determined by the average
correlation energy $\epsilon_c(\Lambda)$,

\begin{equation}
\epsilon_c(\Lambda)=\frac{1}{2\pi} \int dk f(k)
2\sin k~ \theta(\sin k-\Lambda) ,\label{eq18}
\end{equation}

\noindent
through a Boltzmann weight: $w(\Lambda)=\exp(-\epsilon_c(\Lambda)/T)$.

Finally $D(T)$ is given by:
\begin{equation}
D=
\frac{1}{2\pi Z_{\Lambda}}
\int_{-\infty}^{+\infty} d\Lambda g(\Lambda) w(\Lambda)
\frac{1}{2\pi} \int dk f(k) D(\Lambda,k) ,\label{eq19}
\end{equation}

\noindent
with $g(\Lambda)=\partial P/\partial\Lambda$ determined
from Eq.(\ref{eq17}):

\begin{equation}
\frac{\partial P}{\partial \Lambda}=\frac{1}{2\pi}\int dk f(k)
\frac{\partial \theta(\sin k-\Lambda)}{\partial \Lambda} \label{eq20}
\end{equation}

\begin{equation}
Z_{\Lambda}=\frac{1}{2\pi}\int d\Lambda g(\Lambda) w(\Lambda) ,
\label{eq21}
\end{equation}
\noindent
(in the total partition function $Z_{\Lambda}$ we add the
contribution from both the regular and bound states).

\begin{equation}
D(\Lambda,k)=\frac{1}{2}2\sin k
( \frac{\partial^2 \theta}{\partial \Lambda^2}
(\frac{\partial\Lambda}{\partial\phi})^2+
\frac{\partial \theta}
{\partial\Lambda} \frac{\partial^2\Lambda}{\partial\phi^2} ) .
\label{eq22}
\end{equation}

By successive differentiation of Eq.(\ref{eq17}) we can determine
$\partial \Lambda/\partial\phi=1/g(\Lambda)$ and
$\partial^2 \Lambda/\partial \phi^2$.

Evaluating the expressions (\ref{eq19}-\ref{eq22}) we obtain
the results presented in Fig.2, 3 in very good agreement
with the numerical results, providing support for our approach\cite{fut}.
We also verified that the agreement remains (within a couple of percent)
for other values of the interaction (e.g. $U/t=8$).

{\it Comments:}
{}From our formulation of $D(T)$ as a thermal average over
level curvatures, we can try to understand the difference in behavior
between integrable and nonintegrable systems.
In nonintegrable cases the level repulsion prevents level crossings, i.e.
crossings are statistically negligible for macroscopic systems. Then
each level $\epsilon_n(\phi)$ fluctuates only on the scale $\Delta \epsilon
\propto 1/{\cal N}(E)$, where ${\cal N}(E)$ is the many-body density of states.
Therefore the curvature $D_n$ averaged over $\phi$ or over different
$k$ vectors in the thermodynamic limit should vanish.
On the other hand, in integrable
systems levels in general cross, so fluctuations of $\epsilon_n(\phi)$
do not necessarily vanish for $L\to \infty$.
Hence there is no restriction on the average
$D(T)$, except that $D>0$. The difference between both cases is intimately
related to level statistics.

This connection between integrability and finite temperature
charge transport born out of this model
calculation, we can conjecture to hold true for other quantum (as
well as classical) integrable systems.
We can trace it to the existence of a macroscopic number of conservation
laws.
It is plausible that with respect
to transport one should distinguish two types of models within the class of
one-dimensional (1D) correlated systems where
some solvable models\cite{Emer} are available:

\noindent a) In few (mostly solvable) 1D models the
current is a conserved quantity. This is generically the case for models
without Umklapp scattering, e.g. the Luttinger model,
1D Bose gas etc, but
also for the $U=\infty$ Hubbard model \cite{Brin}. In these cases one
expects at $T>0$ ideal conductance of the system characterized by
$\rho(T)=0$ or $D(T)>0$.

\noindent b) Nontrivial answers are expected for models with Umklapp
scattering, e.g. for the 1D Hubbard model, the $t-V$ model etc.
If our conjecture is correct we expect integrable models as
the Hubbard, $t-V$ or supersymmetric
$t-J (J=2t)$ models to behave as ideal conductors at finite temperatures.
We can then argue that even 1D
nonintegrable models as the $U-V$ model
(with longer range interactions), as they are characterized by the
integrable Luttinger liquid Hamiltonian at low energies, should behave as
nearly ideal conductors at low T; the $\delta$-peak then would broaden to
a narrow Drude peak of weight $D(T)$\cite{Giam}.

We should stress that
our study is based on the Kubo linear response theory\cite{Kubo} whose
applicability in the context of Luttinger liquids
has recently been debated\cite{pwa}.
Further work is necessary on other systems to lend further
support for these ideas.

\acknowledgments
We would like to thank H. Schulz, H. Kunz and U.Eckern for useful discussions.
This work was supported by the Swiss National Fond Grants No.
20-39528.93, the University of Geneva, University of Fribourg
and by the Ministry of Science and Technology of Slovenia.

\begin{figure}

\caption{$D(T)/D_0$ as a function of $t_h/t$ for $U/t=2$,
different temperatures $T$ and system sizes $L$:
$( -~\cdot~- )~ L=6$,
$( -~\cdot~\cdot~- )~ L=10$,
$( -~\cdot~\cdot~\cdot~- )~ L=14$.}
\label{one}

\caption{a)lower and left axes: scaling of $D(T=2t)/D_0$ with system
size $1/L$ (circles) for $t_h=t$, $U/t=2$.
The $\odot$ indicates the Bethe ansatz result. The continuous
line is a linear fit for $L=10,14,18$ sites.
b) upper and right axes: scaling of $\ln D(T=2t)/D_0$ with $L^2$ (dots)
for $U/t=2$ and $t_h=0.5, 1.2, 2.0t$}
\label{two}

\caption{$D(T)/D_0$ as a function of $T$ for $U/t=2$. Numerical results
are indicated by
$( -~\cdot~- )~ L=6$,
$( -~\cdot~\cdot~- )~ L=10$,
$( -~\cdot~\cdot~\cdot~- )~ L=14$
and $t_h=0.5t$; by points for $t_h=t$ and $L=18$.
The continuous line is the Bethe ansatz result and the dashed line
$D(T)/D_0$ for a free particle.}
\label{three}

\end{figure}


\begin{references}
\bibitem{Kohn} W. Kohn, Phys. Rev. {\bf 133},171 (1964).
\bibitem{Suth} B.S. Shastry, B. Sutherland, Phys. Rev. Lett.
{\bf 65}, 243 (1990).
\bibitem{Scal} D.J. Scalapino, S.R. White, S.C. Zhang, Phys. Rev. Lett.
{\bf 68}, 2830 (1992); Phys. Rev. {\bf B47}, 7995 (1993).
\bibitem{Kubo} R. Kubo, J. Phys. Soc. Jpn., {\bf 12}, 570 (1957).
\bibitem{Thoul} D.J. Thouless, Phys. Rev. Lett. {\bf 39}, 1167 (1977).
\bibitem{w} M. Wilkinson, J. Phys. {\bf A21}, 4021 (1988);
B. D. Simons, B.L. Altshuler, Phys. Rev. Lett. {\bf 70}, 4063 (1993).
\bibitem{Bell} G. Montambaux, D. Poilblanc, J. Bellisard, C. Sire,
Phys. Rev. Lett. {\bf 70}, 497 (1993).
\bibitem{DiSZ} M.Di Stasio, X. Zotos, preprint.
\bibitem{mald} P.F. Maldague, Phys. Rev. {\bf B16}, 2437 (1977);
D. Baeriswyl, C. Gros and T.M. Rice, Phys. Rev. {\bf B35}, 8391 (1987).
\bibitem{peter} an alternative method, using the Lanczos technique, has
recently been proposed, J. Jakli\v c and P. Prelov\v sek Phys. Rev.
{\bf B49}, 5065 (1994).
\bibitem{mcg} J.B. McGuire, J. Math. Phys. {\bf 6}, 432 (1965);
J. Math. Phys. {\bf 7}, 123 (1966).
\bibitem{xz} X. Zotos, F. Pelzer, Phys. Rev. {\bf B37}, 5045 (1988)
\bibitem{cz} H. Castella, X. Zotos, Phys. Rev. {\bf B47}, 16186 (1993)
\bibitem{lw} E.H. Lieb and F.Y. Wu, Phys. Rev. Lett. {\bf 20}, 1445 (1968).
\bibitem{fut} A complete analysis will be presented elsewhere.
\bibitem{Emer} see e.g. V. Emery, in {\it Highly Conducting One-Dimensional
Solids}, ed. J.T. Devreese et al. (Plenum Press, 1979), p.247.
\bibitem{Brin} B. Brinkman and T.M. Rice, Phys. Rev. {\bf B2}, 6880 (1970).
\bibitem{Giam} T. Giamarchi, Phys. Rev. {\bf B44}, 2905 (1991).
\bibitem{pwa} M. Ogata and P.W. Anderson, Phys. Rev. Lett. {\bf 70},
3087 (1993).

\end{references}
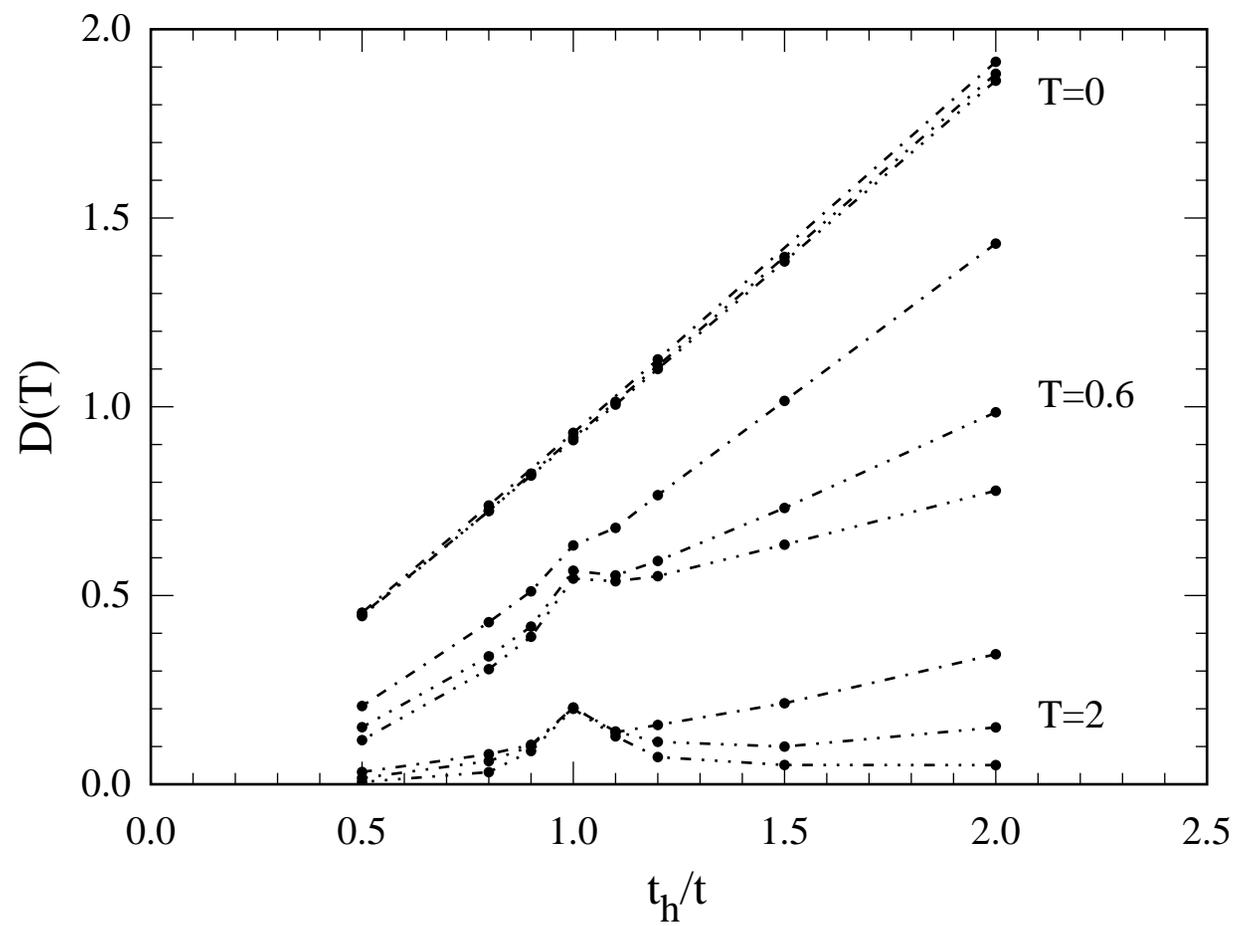
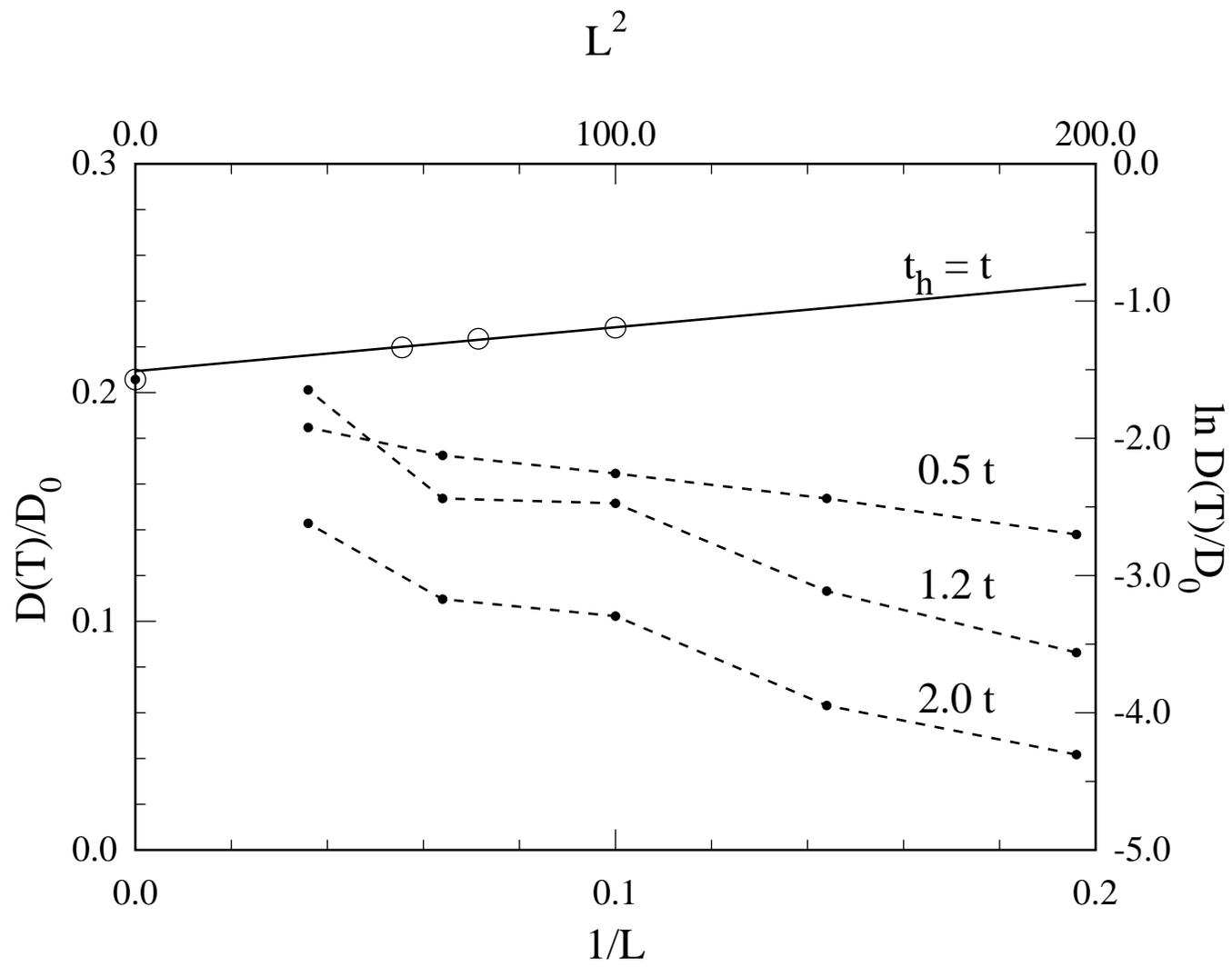
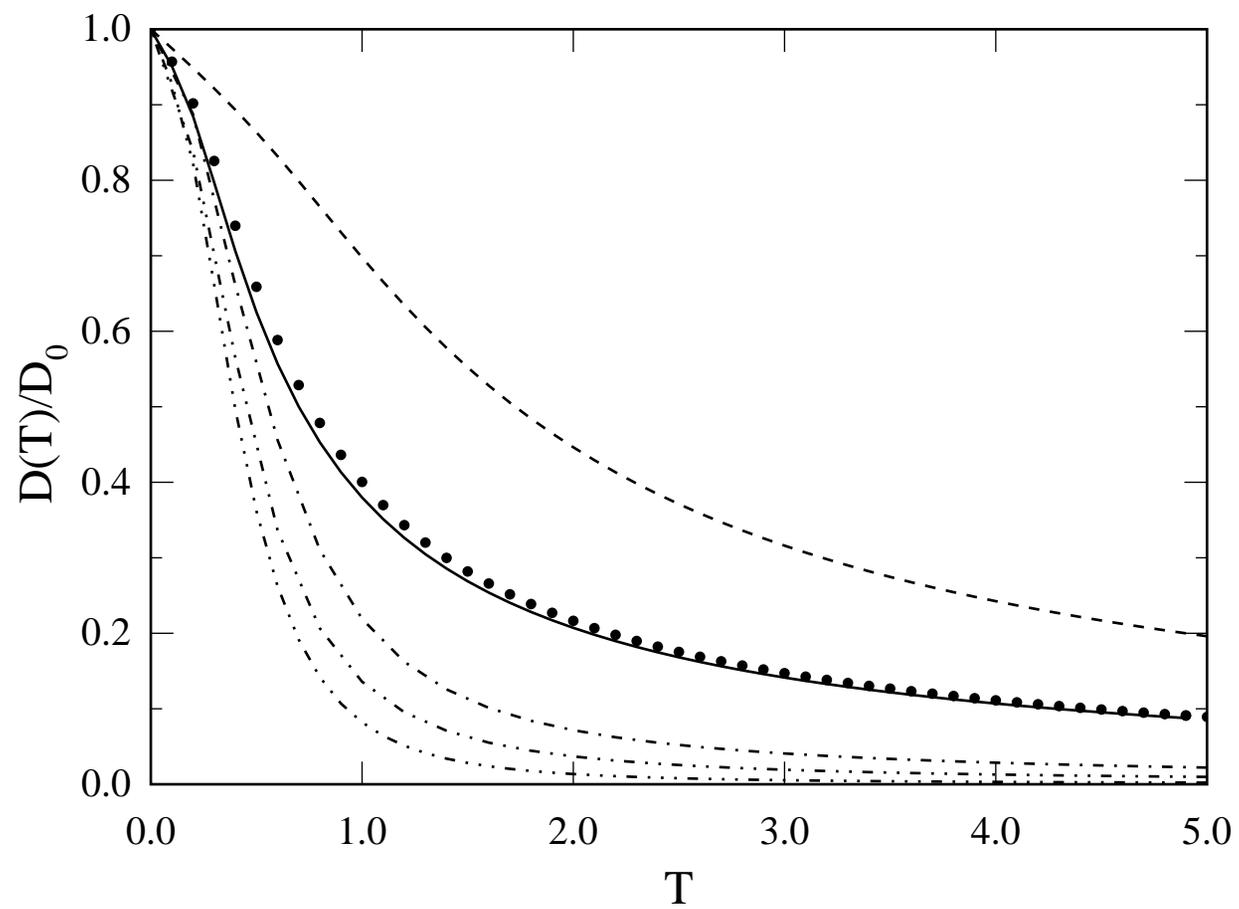
\end{document}